\documentclass[aps,prd,onecolumn,groupedaddress,showpacs,nofootinbib,amssymb]{revtex4}
\usepackage[dvips]{graphicx,color}
\usepackage{amssymb}
\usepackage{amsmath}
\usepackage{graphicx}
\usepackage{amsfonts}
\usepackage{bm}

\newcommand{\be}{\begin{equation}}
\newcommand{\ee}{\end{equation}}
\newcommand{\bea}{\begin{eqnarray}}
\newcommand{\eea}{\end{eqnarray}}
\newcommand{\beaa}{\begin{eqnarray*}}
\newcommand{\eeaa}{\end{eqnarray*}}

\newcommand{\nn}{\nonumber \\}
\newcommand{\e}{\mathrm{e}}

\begin{document}

\title{Aspects of Late-time Evolution in Mimetic $F(R)$ Gravity}
\author{V.~K.~Oikonomou,$^{1,2}$\,\thanks{v.k.oikonomou1979@gmail.com}}
\affiliation{
$^{1)}$ Tomsk State Pedagogical University, 634061 Tomsk, Russia\\
$^{2)}$ Laboratory for Theoretical Cosmology, Tomsk State University of Control Systems
and Radioelectronics (TUSUR), 634050 Tomsk, Russia\\
}

\begin{abstract}
We demonstrate how to describe in an unified way early and late-time acceleration in the context of mimetic $F(R)$ gravity. As we show, an exponential $F(R)$ gravity model has appealing features, with regard to unification, and we perform an analysis of the late-time evolution. The resulting picture is interesting since in the mimetic case, certain pathologies of some ordinary $F(R)$ models are remedied in a consistent way, owing to the presence of the mimetic potential and the Lagrange multiplier. We quantify the late-time evolution analysis by studying the scaled dark energy density, the dark energy equation of state and the total effective equation of state, and as we show the late-time evolution is crucially affected by the functional form of the $F(R)$ gravity. It is intriguing that the most appealing case corresponds to the exponential $F(R)$ gravity which unifies late and early-time acceleration. Finally, we study the behavior of the effective gravitational constant and the growth factor, and as we show, significant differences between the mimetic and ordinary $F(R)$ exponential model are spotted in the growth factor.
\end{abstract}


\maketitle

\section{Introduction}

One of the most unexpected cosmic observations was the late-time acceleration of our Universe in the late 90's \cite{riess}. This observation initiated a research stream with aim to model in a consistent way this late-time acceleration. The ultimate goal of the current research in cosmology is to describe in a unified way the late and early-time acceleration and modified gravity in its various forms can harbor quite successful unified descriptions, see for example the reviews \cite{reviews1,reviews1a,reviews2,reviews3,reviews4,reviews5,rev6,rev7}. For similar approaches in the context of $F(R)$ gravity, see \cite{Appleby:2007vb,Nojiri:2007as,Nojiri:2007cq,Cognola:2007zu}. In this paper we shall investigate how the unification of late and early-time acceleration eras can be achieved in the context of mimetic $F(R)$ gravity. Mimetic $F(R)$ gravity was initially introduced in Ref. \cite{mukhanov1}, and the main feature of this theory was that it resulted to a dark matter component in the gravitational equations, without the need of a perfect fluid, with effective equation of state parameter being equal to zero, that is, $w=0$. This extra dark mater component appeared in the equations of motion due to the inherent conformal degree of freedom of the metric, see \cite{mukhanov1} for more details. The mimetic gravity framework was extensively studied later on in Refs. \cite{mukhanov2,Golovnev1,Golovnev2,Golovnev3,Golovnev4,Golovnev5,Golovnev6,Golovnev7,NO2,mimetic1,mimetic2,mimetic3,mimetic4,mimetic5,mimetic8,mimetic9,mimetic10,mimetic11,mimetic12}, and was generalized in the context of $F(R)$ gravity in Ref. \cite{NO2}. As it was shown in Ref. \cite{NO2}, the mimetic constraint can be produced by introducing a Lagrange multiplier in the action \cite{CMO1,CMO3}, and also the existence of a potential is allowed, which we will call mimetic potential for brevity. Actually the presence of the mimetic potential and of the Lagrange multiplier makes possible the realization of various cosmologies and also makes possible to describe a realistic late-time behavior, for which the pathologies of some ordinary $F(R)$ gravity models are remedied in a self-consistent way.

Particularly, in this paper we shall investigate the certain aspects of the late-time evolution of some mimetic $F(R)$ gravity models. It is conceivable that the choice of the $F(R)$ gravity plays a crucial role for the realization of a viable cosmology, so we shall investigate how the functional form of the $F(R)$ gravity affects the resulting cosmology. In addition, the choice of the $F(R)$ gravity affects the potential and the Lagrange multiplier, so the resulting physical picture crucially depends on the choice of the $F(R)$ gravity. In principle the presence of the potential and Lagrange multiplier offers much freedom for realizing various cosmic scenarios, however there are certain theoretical aspects that crucially depend on the choice of the $F(R)$ gravity. Particularly the dark energy oscillations \cite{staro,exp0,exp1,exp2,exp2a,exp3a,exp4}, which can be generated at the end of the matter domination era when the redshift is nearly $z=3$, can cause complications in some ordinary $F(R)$ gravity models. As we show, the dark energy oscillations issue can also exist in the mimetic $F(R)$ gravity case, and it crucially depends on the choice of the $F(R)$ gravity. The novel feature of mimetic $F(R)$ gravity is that the form of the $F(R)$ gravity needs not to be modified by hand by adding power-law corrections, as was the case in some ordinary $F(R)$ gravity models in order to reduce the amplitude of the dark energy oscillations. Also a study of the dark energy equation of state parameter and also of the total effective equation of state parameter reveals that the mimetic $F(R)$ gravity case can yield very interesting results related to the late-time era, for some specific models. Finally, our study of the late-time evolution will include an investigation of the effective gravitational constant and of the growth index for some mimetic $F(R)$ models. The most important outcomes of our work are that firstly a unified description of late and early-time acceleration can be achieved and secondly, cosmologies with viable and realistic late-time era can be consistently realized by some mimetic $F(R)$ models.

This paper is organized as follows: In section II we present in brief the essential features of mimetic $F(R)$ gravity and also we thoroughly investigate the late-time evolution of two viable $F(R)$ gravity models in the context of mimetic gravity. As we show, the choice of the model crucially affects the late-time evolution. In section III, we study the behavior of the effective gravitational constant and of the growth index for certain mimetic $F(R)$ models and we compare with ordinary $F(R)$ gravity. Finally the conclusions follow in the end of the paper.

\section{Late-time Evolution in Exponential and Power Law Mimetic $F(R)$ Gravity}

The formalism of mimetic $F(R)$ gravity was introduced in \cite{NO2}, where a mimetic potential and a Lagrange multiplier was used, with the corresponding gravitational action being,
\be
\label{MF1}
S = \int \sqrt{-g} \left\{ \frac{F(R)}{2\kappa^2} + \lambda \left( \partial_\mu \phi \partial^\mu \phi
+ 1 \right) - V(\phi) \right\} + S_\mathrm{matter} \left( g_{\mu\nu}, \Psi \right)\, ,
\ee
where $S_\mathrm{matter}$ stands for the action of the matter fluids which are present. By varying the action (\ref{MF1}) with respect to the physical metric $g_{\mu \nu}$ metric, we get,
\begin{align}
\label{MG14}
0= & \frac{1}{2\kappa^2} \left\{ \frac{1}{2} g_{\mu\nu} F(R) - R_{\mu\nu} F' (R)
+ \nabla_\mu \nabla_\nu F'(R) - g_{\mu\nu} \nabla^2 F'(R) \right\}  \nn
& + \frac{1}{2} g_{\mu\nu} \left\{ \lambda \left( \partial_\mu \phi \partial^\mu \phi + 1 \right)
 - V(\phi) \right\} - \lambda \partial_\mu \phi \partial_\nu \phi +\frac{1}{2} T_{\mu\nu} \, .
\end{align}
In addition, by varying the action with respect to the scalar field $\phi$ and with respect to the Lagrange multiplier, we obtain the following equations,
\begin{align}
\label{MG15}
0 = - 2 \nabla^\mu \left( \lambda \nabla_\mu \phi \right) - V'(\phi)\, , \\
\label{MG16}
0=\partial_\mu \phi \partial^\mu \phi + 1\, .
\end{align}
Assuming a flat Friedmann-Robertson-Walker metric (FRW),
\be
\label{MG17}
ds^2 = - dt^2 + a(t)^2 \sum_{i=1,2,3} \left( dx^i \right)^2 \, ,
\ee
and also that $\phi$ depends only on the cosmic time $t$, the mimetic constraint, namely Eq. (\ref{MG16}), yields the following solution,
\be
\label{MG18}
\phi = t \, ,
\ee
therefore, the Eq.~(\ref{MG15}) takes the following form,
\be
\label{MG19}
0 = 2 \dot \lambda + 6 H \lambda - V'(t)\, ,
\ee
where $H= \dot a /a$ is the Hubble rate. Moreover, the $(t,t)$ and $(i,j)$ $\left( i,j=1,2,3 \right)$ components of the equations (\ref{MG14}) take the following form:
\begin{align}
\label{MG20}
0= & \frac{1}{2\kappa^2} \left\{ - \frac{1}{2} F(R) + 3 \left( H^2 + \dot H \right) F'(R)
 - 18 \left( 4 H^2 \dot H+ H \ddot H \right) F''(R) \right\} + \frac{1}{2} V(t)
 - \lambda + \frac{1}{2} \rho \, , \\
\label{MG21}
0= & \frac{1}{2\kappa^2} \left\{ \frac{1}{2} F(R) - \left( \dot H + 3 H^2 \right) F'(R)
+ 6 \left( 8 H^2 \dot H + 4 {\dot H}^2 + 6 H \ddot H + \dddot H \right) F''(R)
+ 36 \left( 4 H \dot H + \ddot H \right)^2 F''' (R) \right\} \nn
& - \frac{1}{2} V(t) + \frac{1}{2} p \, ,
\end{align}
with $\rho$ and $p$ being the energy density and the pressure of the matter fluids present, and in addition the Ricci scalar curvature is equal to $R=12H^2 + 6 \dot H$. In principle, for a given potential, Lagrange multiplier and $F(R)$ gravity, any arbitrary cosmological evolution can be realized from mimetic $F(R)$ gravity. Consider the model with the following scale factor,
\be
\label{MG23}
a(t) = a_0 \left( \e^{-\frac{t}{t_0}} + 1 \right)^{- t_0 H_I} \e^{H_L t}\, ,
\ee
with $a_0$ being an arbitrary constant. The corresponding Hubble rate is,
\be
\label{MG22}
H = \frac{H_I \e^{- \frac{t}{t_0}}}{\e^{-\frac{t}{t_0}} + 1} + H_L \, .
\ee
By choosing the parameters $H_I$ and $H_L$ as $H_I\gg H_L>0$ in the limit $t\to -\infty$, the Hubble rate tends to $H\to H_I$ and in the late-time regime $t\to +\infty$, it tends to $H\to H_L$. Hence this cosmological evolution at early times describes an inflationary de Sitter regime with $a(t)\sim e^{H_I t}$ and also at late times it describes an accelerating expansion with $H\sim H_L$. This kind of evolution can be generated by exponential $F(R)$ gravity \cite{exp0,exp1,exp2,exp2a,exp3a,exp4} of the form,
\begin{equation}\label{expmodnocurv}
F(R)=R-2\Lambda \left ( 1-e^{\frac{R}{\beta \Lambda}} \right )\, ,
\end{equation}
where $\Lambda$ is the cosmological constant corresponding to present time and the parameter $\beta$ is a positive free parameter of the order $\mathcal{O}(1)$. In the following sections we shall investigate the implications of the unification cosmology (\ref{MG23}) for a mimetic $F(R)$ gravity with potential and Lagrange multiplier. Particularly, we shall choose the $F(R)$ gravity to be of the form (\ref{expmodnocurv}) and we shall find the corresponding mimetic potential which corresponds at late times. Then for the resulting potential, we shall investigate the qualitative features of the cosmological evolution at late times and we shall compare the results to the standard $F(R)$ case and also to other mimetic $F(R)$ gravities.

As we already discussed in the introduction, the novel feature of the mimetic formalism is that dark matter appears in the gravitational equations as a consequence of the internal conformal degree of freedom of the metric. In this section we shall present another interesting feature of the mimetic formalism, in the absence of cold dark matter. Particularly we shall assume that the universe is described by a mimetic $F(R)$ gravity in the presence of a single matter fluid which describes radiation. We are interested in finding the late-time behavior of the cosmic evolution (\ref{MG23}), and we shall compare the resulting picture to the standard $F(R)$ gravity results. We focus on three physical quantities and their late-time behavior, namely, the total effective equation of state, the dark energy density and the scaled dark energy, which we define later on. As we show these quantities behave in a different way for some mimetic $F(R)$ models, in comparison to the corresponding ordinary $F(R)$ models.

We shall use two viable $F(R)$ gravity models, namely the one appearing in Eq. (\ref{expmodnocurv}) and also just for expositional purposes, we shall also take into account the following model,
\begin{equation}\label{powerlaw}
F(R)=R-\mu R^{n}\, ,
\end{equation}
with $\mu >0$ and $0<n<1$. The model (\ref{powerlaw}) does not have the appealing properties of the model (\ref{expmodnocurv}), but as we already mentioned, we study this model in order to have a complete overview of the late-time behavior and how this is affected by the models.

A major difference between the mimetic and non-mimetic $F(R)$ description, is that in the ordinary $F(R)$ case, for some models extra correction terms are needed in order to avoid the problem of dark energy oscillations \cite{exp0,exp1,exp4}. This issue occurs owing to the fact that the higher order derivatives of the dark energy effective energy density take large values, which in turn affects the dark energy equation of state parameter $\omega_{DE}$ \cite{exp0}. A remedy for this issue was proposed in Ref. \cite{exp1}, which is a modification of the model (\ref{MG23}), and the $F(R)$ gravity is,
\begin{equation}\label{expmodnocurvcorr}
F(R)=R-2\Lambda \left ( 1-e^{\frac{R}{\beta \Lambda}} \right )-\tilde{\gamma}\Lambda \left (\frac{R}{3\tilde{m}^2} \right )^{1/3}
\end{equation}
where $\Lambda=7.93\tilde{m}^2$ and $\tilde{\gamma}=1/1000$. The addition of the last term in (\ref{expmodnocurvcorr}) plays a crucial role, since this term stabilizes the dark energy oscillations near the end of the matter domination era and during the late-time acceleration era. A major difference between the mimetic and ordinary $F(R)$ gravity is that in the case of mimetic $F(R)$, the extra curvature term in (\ref{expmodnocurvcorr}) is not actually needed and the dark energy oscillations are dumped for the exponential model (\ref{MG23}), a feature which is particularly interesting. This important difference occurs owing to the existence of the mimetic potential and of the Lagrange multiplier as we evince later on in this section.

Before we proceed to the comparison, it is worth  introducing new variables that are constructed to measure the dark energy oscillations. The rest of the physical quantities, namely the dark energy density $\Omega_{DE}(z)$ and the effective equation of state parameter $\omega_{eff}(z)$, can be expressed in terms of these new variables. Also notice that we made use of the variable $z$, the redshift, so all the physical quantities will be expressed in terms of the redshift $z$.

In order to describe in an optimal way the dark energy oscillations, we recast the FRW equation (\ref{MG20}) as follows,
\begin{align}\label{eq:flrw}
& 3F'H^2=\frac{\rho}{2}+\frac{V(t)-2\lambda(t)}{2}+\frac{1}{2}(F'R-F)-3H\dot{F'}\, ,
\end{align}
where $\rho$ is the energy density of the perfect fluids present, which for the mimetic case will be radiation, so,
\begin{equation}\label{totalmattenergdensmf1}
\rho=\rho_{r}^{(0)}a^{-4}\, ,
\end{equation}
while for the ordinary $F(R)$ gravity case, the total energy density is,
 \begin{equation}\label{totalmattenergdensmf}
\rho_{matt}=\rho_m^{(0)}a^{-3}+\rho_{r}^{(0)}a^{-4}\, .
\end{equation}
The equation (\ref{eq:flrw}) can be written as follows,
\begin{equation}\label{eq:modifiedeinsteineqns2}
  H^2-(F'-1)\left (H\frac{\mathrm{d}H}{\mathrm{d}\ln{a}}+H^2 \right )+\frac{1}{6}(F-R)+H^2F''\frac{\mathrm{d}R}{\mathrm{d}\ln{a}}-\frac{V(t)-2\lambda(t)}{3}=\frac{\rho_{matt}}{3},
\end{equation}
where $R$ is the Ricci scalar, which we write as follows,
\begin{equation}\label{eq:ricciscal2}
  R=12H^2+6H\frac{\mathrm{d}H}{\mathrm{d}\ln{a}}.
\end{equation}
We introduce the following function of the redshift,
\begin{equation}\label{potlagra}
  \mathcal{Q}(a(z))=V(a(z))-2\lambda (a(z))\, ,
\end{equation}
which depends on the mimetic potential and on the Lagrange multiplier. In order to capture the effects of the dark energy oscillations and perfectly quantify the late-time evolution, we introduce the following variables,
\begin{subequations}
  \begin{align}
    y_H&\equiv\frac{\rho_{DE}}{\rho_m^{(0)}}=\frac{H^2}{\tilde{m}^2}-\mathcal{Q}(a)-\chi a^{-4}\, ,\label{eq:yH}\\
    y_R&\equiv\frac{R}{\tilde{m}^2}-\frac{\mathrm{d}\mathcal{Q}(a)}{\mathrm{d}\ln{a}}\, ,\label{eq:yR}
  \end{align}
\end{subequations}
where $\rho_{DE}$ denotes the total energy density of dark energy, and also $\rho_m^{(0)}$ and $\rho_r^{(0)}=\chi\rho_m^{(0)}$ stand for the present values of the mass-energy densities corresponding to cold dark matter and radiation respectively. Moreover, the parameter $\chi$ is the ratio $\chi=\rho_r^{(0)}/\rho_m^{(0)}$. By diving Eqs. (\ref{eq:modifiedeinsteineqns2}) by $\tilde{m}^2$ and also by using the following equation,
\begin{equation}
  \frac{H^2}{\tilde{m}^2}-\frac{\rho_M}{2\tilde{m}^2}=\frac{H^2}{\tilde{m}^2}-\mathcal{Q}(a)-\chi a^{-4}=y_H
\end{equation}
and also having in mind Eq. (\ref{eq:yH}), we can solve the resulting equation with respect to $\frac{1}{\tilde{m}^2}\frac{dR}{d\ln{a}}$. Then differentiating the expression in Eq. (\ref{eq:yR}) with respect to $\ln{a}$ we obtain,
\begin{equation}\label{eq:dyR}
  \frac{\mathrm{d}y_R}{\mathrm{d}\ln{a}}=-\frac{\mathrm{d}^2\mathcal{Q}(a)}{\mathrm{d}\ln{a}^2}+\left[-y_H+(F'-1)(\frac{H}{\tilde{m}^2}\frac{\mathrm{d}H}{\mathrm{d}\ln{a}}
    +\frac{H^2}{\tilde{m}^2})-\frac{1}{6\tilde{m}^2}(F-R)\right]\frac{1}{H^2F''}\, .
\end{equation}
Accordingly, by using Eq. (\ref{eq:yH}), differentiating Eq. (\ref{eq:yH}) with respect to $\ln{a}$ and also by using Eq. (\ref{eq:dyR}), we obtain,
\begin{eqnarray}\label{eq:dyR2}
  \frac{\mathrm{d}y_R}{\mathrm{d}\ln{a}}&=&-\frac{\mathrm{d}^2\mathcal{Q}(a)}{\mathrm{d}\ln{a}^2}+
    \left[-y_H+(F'-1)\left(\frac{1}{2}\frac{\mathrm{d}y_H}{\mathrm{d}\ln{a}}+\frac{1}{2}\frac{\mathrm{d}\mathcal{Q}(a)}{\mathrm{d}\ln{a}}
    +y_H+\mathcal{Q}(a)-\chi a^{-4}\right)\right.\nonumber\\
    &&\left.-\frac{1}{6\tilde{m}^2}(F-R)\right]\frac{1}{\tilde{m}^2F''(y_H+\mathcal{Q}(a)+\chi a^{-4})}.
\end{eqnarray}
Moreover, upon differentiation of Eq. (\ref{eq:dyR2}) with respect to $\ln{a}$ we obtain,
\begin{equation}\label{eq:dyH}
\frac{\mathrm{d}y_H}{\mathrm{d}\ln{a}}=\frac{2H}{\tilde{m}^2}\frac{\mathrm{d}H}{\mathrm{d}\ln{a}}-\frac{\mathrm{d}\mathcal{Q}(a)}{\mathrm{d}\ln{a}}+4\chi a^{-4}\, ,
\end{equation}
and combining Eqs. (\ref{eq:ricciscal2}) and (\ref{eq:yH}), the Eq. (\ref{eq:dyH}) can be cast as follows,
\begin{eqnarray}\label{eq:dyH2}
  \frac{\mathrm{d}y_H}{\mathrm{d}\ln{a}}&=&\frac{R}{3\tilde{m}^2}-\frac{\mathrm{d}\mathcal{Q}(a)}{\mathrm{d}\ln{a}}-4y_H-4\mathcal{Q}(a)\nonumber\\
    &=&\frac{y_R}{3}-4y_H-\frac{2\mathrm{d}\mathcal{Q}(a)}{3\mathrm{d}\ln{a}}-4\mathcal{Q}(a)\, .
\end{eqnarray}
Then, by writing the Ricci scalar in the following way,
\begin{equation}
  R=3\tilde{m}^2\left(4y_H+4\mathcal{Q}(a)+\frac{\mathrm{d}y_H}{\mathrm{d}\ln{a}}+\frac{\mathrm{d}\mathcal{Q}(a)}{\mathrm{d}\ln{a}}\right)\, ,
\end{equation}
by differentiating Eq. (\ref{eq:dyH2}) with respect to $\ln{a}$ and also by using Eq. (\ref{eq:dyR2}), we obtain,
\begin{eqnarray}\label{eq:FRform}
  &&\frac{\mathrm{d}^2y_H}{\mathrm{d}\ln{a}^2}+\left(4+\frac{1-F'}{6\tilde{m}^2F''(y_H+\mathcal{Q}(a)+
    \chi a^{-4})}\right)\frac{\mathrm{d}y_H}{\mathrm{d}\ln{a}}
    +\left(\frac{2-F'}{3\tilde{m}^2F''(y_H+\mathcal{Q}(a)+\chi a^{-4})}\right)y_H
  \nonumber\\
  &&+\left(\frac{\mathrm{d}^2\mathcal{Q}(a)}{\mathrm{d}\ln{a}^2}+4\frac{\mathrm{d}\mathcal{Q}(a)}{\mathrm{d}\ln{a}}
    +\frac{(1-F')\left(3\frac{\mathrm{d}\mathcal{Q}(a)}{\mathrm{d}\ln{a}}+6\mathcal{Q}(a)-6\chi a^{-4}\right)+\frac{F-R}{\tilde{m}^2}}
      {18\tilde{m}^2F'' (y_H+\mathcal{Q}(a)+\chi a^{-4})}\right)=0
\end{eqnarray}
Finally, by expressing all the physical quantities in Eq. (\ref{eq:FRform}) as functions of the redshift $z$ according to the following two rules,
\begin{subequations}
  \begin{align}
    \frac{\mathrm{d}}{\mathrm{d}\ln{a}}&=-(z+1)\frac{\mathrm{d}}{\mathrm{d}z},\\
    \frac{\mathrm{d}^2}{\mathrm{d}\ln{a}^2}&=(z+1)\frac{\mathrm{d}}{\mathrm{d}z}+(z+1)^2\frac{\mathrm{d}^2}{\mathrm{d}z^2}\, ,
  \end{align}
\end{subequations}
we obtain the equation which quantifies the late-time evolution of the mimetic $F(R)$ gravity, which is,
\begin{eqnarray}\label{mastereqn}
  &&\frac{\mathrm{d}^2y_H}{\mathrm{d}z^2}+
    \frac{1}{(z+1)}\left(-3-\frac{F'(R)}{6\tilde{m}^2F''(R)(y_H+\mathcal{Q}(z)+\chi(z+1)^4)}\right)\frac{\mathrm{d}y_H}{\mathrm{d}z}
  \nonumber\\
  &&+\frac{1}{(z+1)^2}\frac{1-F'(R)}{3\tilde{m}^2F''(R)(y_H+\mathcal{Q}(z)+\chi(z+1)^4)}y_H+\left(\frac{
    \mathrm{d}^2\mathcal{Q}(z)}{\mathrm{d}z^2}-\frac{3}{(z+1)}\frac{\mathrm{d}\mathcal{Q}(z)}{\mathrm{d}z}\right.
  \nonumber\\
  &&\left.-\frac{1}{(z+1)}\frac{F'(R)\left(-(z+1)\frac{\mathrm{d}\mathcal{Q}(z)}{
    \mathrm{d}z}+2\mathcal{Q}(z)-2\chi(z+1)^4\right)+\frac{F}{3\tilde{m}^2}}{6\tilde{m}^2F''(R)(y_H-\mathcal{Q}(z)+\chi(z+1)^4)}\right)=0\, .
\end{eqnarray}
By looking the master equation (\ref{mastereqn}), it is obvious that the evolution is strongly affected by the mimetic potential and the Lagrange multiplier which are contained in the function $\mathcal{Q}(z)$. Therefore, in the following we shall specify the exact form of the potential and the Lagrange multiplier, and these will be determined by the evolution (\ref{MG23}) and also by specifying the $F(R)$ gravity. Then, by using the resulting expressions for $V(t)$ and $\lambda (t)$, expressed in terms of the redshift, we will solve numerically the differential equation (\ref{mastereqn}) and we will investigate the behavior of the late-time evolution. Since we are interested at late times, the scale factor (\ref{MG23}) behaves as follows,
\begin{equation}\label{latetimescale}
a(t)\simeq a_0\left( 1-H_I t_0 e^{-t/t_0}\right)e^{H_L t}\, .
\end{equation}
It is conceivable that the scale factor (\ref{latetimescale}) is actually a nearly de Sitter solution since the term $\sim e^{-t/t_0}$ gives very small contribution. However, we need to find the functions $V(t)$ and $\lambda (t)$ at leading order in $t$, and if we use the de Sitter scale factor, the potential and the Lagrange multiplier will be constant, so we use the leading order expression (\ref{latetimescale}).

\subsection{The Exponential $F(R)$ Gravity Case}

At this point we need to specify the functional form of the $F(R)$ gravity, so suppose that the $F(R)$ gravity is the exponential one appearing in Eq. (\ref{expmodnocurv}). Then, the mimetic potential at leading order in the limit $t\to \infty $ reads,
\begin{align}\label{leadingordpotexp}
& V(t)\simeq \frac{6 H_L^2}{\kappa ^2}-\frac{\Lambda }{\kappa ^2}+\frac{e^{-\frac{12 H_L^2}{\beta  \Lambda }} \Lambda }{\kappa ^2}+\frac{144 e^{-\frac{t}{t_0}-\frac{12 H_L^2}{\beta  \Lambda }} H_I H_L^4}{\beta ^2 \kappa ^2 \Lambda }-\frac{36 e^{-\frac{t}{t_0}-\frac{12 H_L^2}{\beta  \Lambda }} H_I H_L^3}{t_0 \beta ^2 \kappa ^2 \Lambda }\\ \notag &
-\frac{720 e^{-\frac{2 t}{t_0}-\frac{12 H_L^2}{\beta  \Lambda }} H_I^2 H_L^3}{t_0^3 \beta ^2 \kappa ^2 \Lambda }+\frac{72 e^{-\frac{2 t}{t_0}-\frac{12 H_L^2}{\beta  \Lambda }} H_I^2 H_L^2}{t_0^4 \beta ^2 \kappa ^2 \Lambda }+\frac{2304 e^{-\frac{2 t}{t_0}-\frac{12 H_L^2}{\beta  \Lambda }} H_I^2 H_L^4}{t_0^2 \beta ^2 \kappa ^2 \Lambda }\\ \notag &
-\frac{2304 e^{-\frac{2 t}{t_0}-\frac{12 H_L^2}{\beta  \Lambda }} H_I^2 H_L^5}{t_0 \beta ^2 \kappa ^2 \Lambda }+\frac{432 e^{-\frac{2 t}{t_0}-\frac{12 H_L^2}{\beta  \Lambda }} H_I^2 H_L^4 t_0}{\beta ^2 \kappa ^2 \Lambda }\, ,
\end{align}
and the Lagrange multiplier $\lambda (t)$ is,
\begin{align}\label{lambdaleading}
& \lambda (t)=-\frac{6 e^{-\frac{12 H_L^2}{\beta  \Lambda }} H_L^2}{\beta  \kappa ^2}+\frac{\Lambda }{2 \kappa ^2}-\frac{e^{-\frac{12 H_L^2}{\beta  \Lambda }} \Lambda }{2 \kappa ^2}-\frac{18 e^{-\frac{t}{t_0}-\frac{12 H_L^2}{\beta  \Lambda }} H_I H_L^3}{t_0 \beta ^2 \kappa ^2 \Lambda }+\frac{72 e^{-\frac{t}{t_0}-\frac{12 H_L^2}{\beta  \Lambda }} H_I H_L^4}{\beta ^2 \kappa ^2 \Lambda }\\ \notag &
-\frac{60 e^{-\frac{2 t}{t_0}-\frac{12 H_L^2}{\beta  \Lambda }} H_I^2 H_L^3}{\beta ^2 \kappa ^2 \Lambda }-\frac{144 e^{-\frac{2 t}{t_0}-\frac{12 H_L^2}{\beta  \Lambda }} H_I^2 H_L^3}{t_0^3 \beta ^2 \kappa ^2 \Lambda }-\frac{576 e^{-\frac{2 t}{t_0}-\frac{12 H_L^2}{\beta  \Lambda }} H_I^2 H_L^4}{t_0^2 \beta ^2 \kappa ^2 \Lambda }\\ \notag &
+\frac{6 e^{-\frac{2 t}{t_0}-\frac{12 H_L^2}{\beta  \Lambda }} H_I^2 H_L^2}{t_0 \beta ^2 \kappa ^2 \Lambda }+\frac{2304 e^{-\frac{2 t}{t_0}-\frac{12 H_L^2}{\beta  \Lambda }} H_I^2 H_L^5}{t_0 \beta ^2 \kappa ^2 \Lambda }+\frac{216 e^{-\frac{2 t}{t_0}-\frac{12 H_L^2}{\beta  \Lambda }} H_I^2 H_L^4 t_0}{\beta ^2 \kappa ^2 \Lambda }\, .
\end{align}
Notice that the expressions (\ref{leadingordpotexp}) and (\ref{lambdaleading}) are identical to the corresponding de Sitter cosmology expressions once the exponentials are erased, with the de Sitter expressions being,
\begin{equation}\label{desitterexpressions}
V(t)=\frac{6 H_L^2}{\kappa ^2}-\frac{\Lambda }{\kappa ^2}+\frac{e^{-\frac{12 H_L^2}{\beta  \Lambda }} \Lambda }{\kappa ^2}\, ,
\end{equation}
\begin{equation}\label{desitter2}
\lambda(t)=-\frac{6 e^{-\frac{12 H_L^2}{\beta  \Lambda }} H_L^2}{\beta  \kappa ^2}+\frac{\Lambda }{2 \kappa ^2}-\frac{e^{-\frac{12 H_L^2}{\beta  \Lambda }} \Lambda }{2 \kappa ^2}\, .
\end{equation}
Also, as we shall see, by appropriately choosing the variables, the mimetic potential vanishes in the limit $t\to \infty$. Then, by assuming that the late-time evolution is the de Sitter one and expressing the potential and the Lagrange multiplier as functions of the redshift, we obtain the final expression for the function $\mathcal{Q}(z)$, which is,
\begin{align}\label{qz1}
& \mathcal{Q}(z)=\frac{6 H_L^2}{\kappa ^2}+\frac{12 e^{-\frac{12 H_L^2}{\beta  \Lambda }} H_L^2}{\beta  \kappa ^2}+\frac{120 e^{-\frac{12 H_L^2}{\beta  \Lambda }} H_I^2 H_L^3 (1+z)^{\frac{2}{H_L t_0}}}{\beta ^2 \kappa ^2 \Lambda }+\frac{72 e^{-\frac{12 H_L^2}{\beta  \Lambda }} H_I^2 H_L^2 (1+z)^{\frac{2}{H_L t_0}}}{t_0^4 \beta ^2 \kappa ^2 \Lambda }\\ \notag & -\frac{432 e^{-\frac{12 H_L^2}{\beta  \Lambda }} H_I^2 H_L^3 (1+z)^{\frac{2}{H_L t_0}}}{t_0^3 \beta ^2 \kappa ^2 \Lambda }+\frac{3456 e^{-\frac{12 H_L^2}{\beta  \Lambda }} H_I^2 H_L^4 (1+z)^{\frac{2}{H_L t_0}}}{t_0^2 \beta ^2 \kappa ^2 \Lambda }-\frac{12 e^{-\frac{12 H_L^2}{\beta  \Lambda }} H_I^2 H_L^2 (1+z)^{\frac{2}{H_L t_0}}}{t_0 \beta ^2 \kappa ^2 \Lambda }\\ \notag &
-\frac{6912 e^{-\frac{12 H_L^2}{\beta  \Lambda }} H_I^2 H_L^5 (1+z)^{\frac{2}{H_L t_0}}}{t_0 \beta ^2 \kappa ^2 \Lambda }-\frac{2 \Lambda }{\kappa ^2}+\frac{2 e^{-\frac{12 H_L^2}{\beta  \Lambda }} \Lambda }{\kappa ^2}\, .
\end{align}
Having the function $\mathcal{Q}(z)$ at hand, we assume that the late-time evolution is not known and so we solve numerically the differential equation (\ref{mastereqn}), by using the following initial conditions,
\begin{equation}\label{initialcond}
y_H(z)\mid_{z=z_{f}}=\frac{\Lambda }{3\tilde{m}^2}\left(1+\frac{z_{f}+1}{1000}\right),{\,}{\,}{\,}y_{H}'(z)\mid_{z=z_{f}}=\frac{\Lambda }{3\tilde{m}^2}\frac{1}{1000}
\end{equation}
with $z_{f}=10$ and $\Omega_M=0.279$. For the mimetic case, the $F(R)$ gravity is that of Eq. (\ref{expmodnocurv}), and we shall compare the resulting evolution to the ordinary $F(R)$ case, with the $F(R)$ gravity in this case being the one of Eq. (\ref{expmodnocurvcorr}). In addition for the numerical analysis we shall use the following values for the parameters, $H_L/H_I=10^{-37}$, $t_0=10^{-33}$sec. By finding the function $y_H(z)$ will enable us to study the dark energy density $\omega_{DE}=P_{DE}/\rho_{DE}$ which is equal to,
\begin{equation}\label{deeqnstateprm}
\omega_{DE}(z)=-1+\frac{1}{3}(z+1)\frac{1}{y_H(z)}\frac{\mathrm{d}y_H(z)}{\mathrm{d}z}\, ,
\end{equation}
and also the total equation of state parameter $\omega_{eff}(z)$, which is equal to,
 \begin{equation}\label{effeqnofstateform}
\omega_{eff}(z)=-1+\frac{2(z+1)}{3H(z)}\frac{\mathrm{d}H(z)}{\mathrm{d}z}\, .
\end{equation}
\begin{figure}[h]
\centering
\includegraphics[width=15pc]{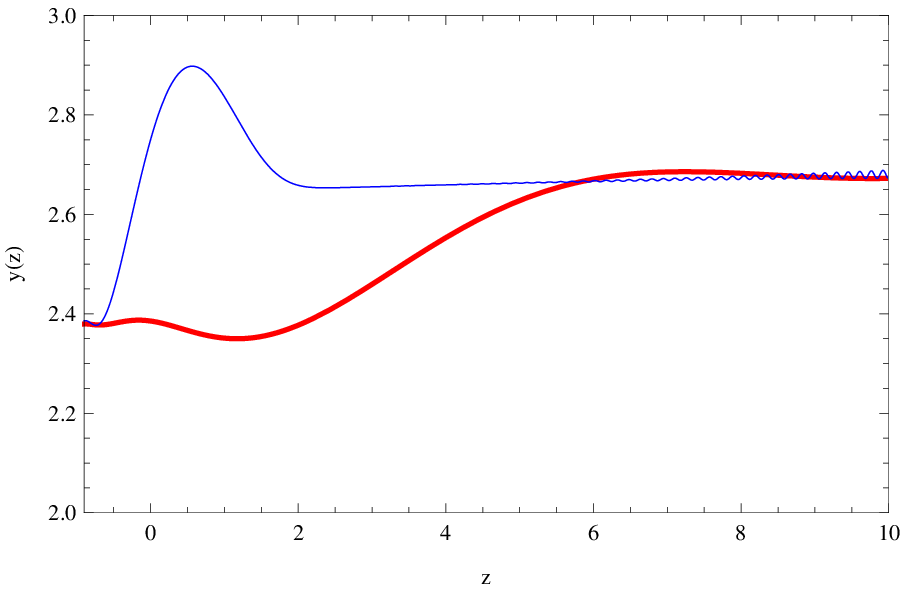}
\includegraphics[width=15pc]{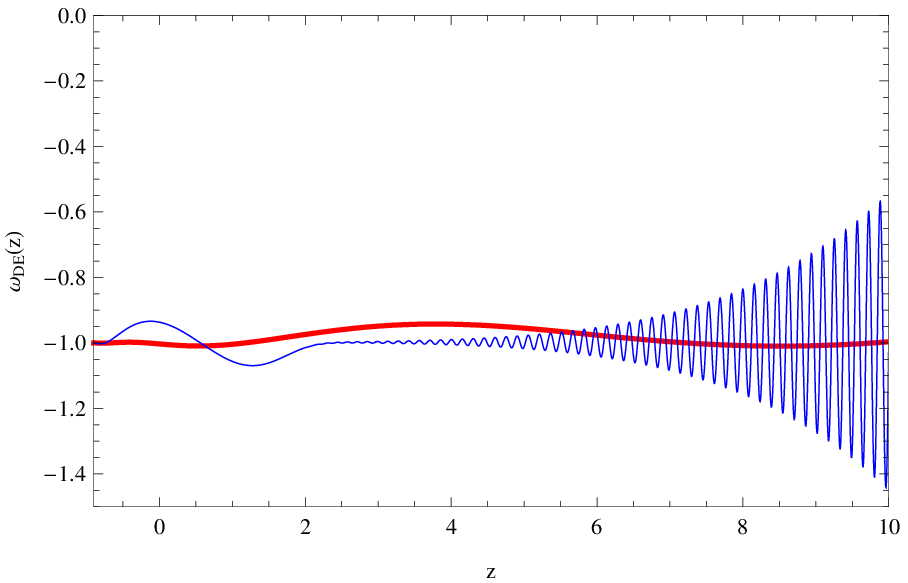}
\includegraphics[width=15pc]{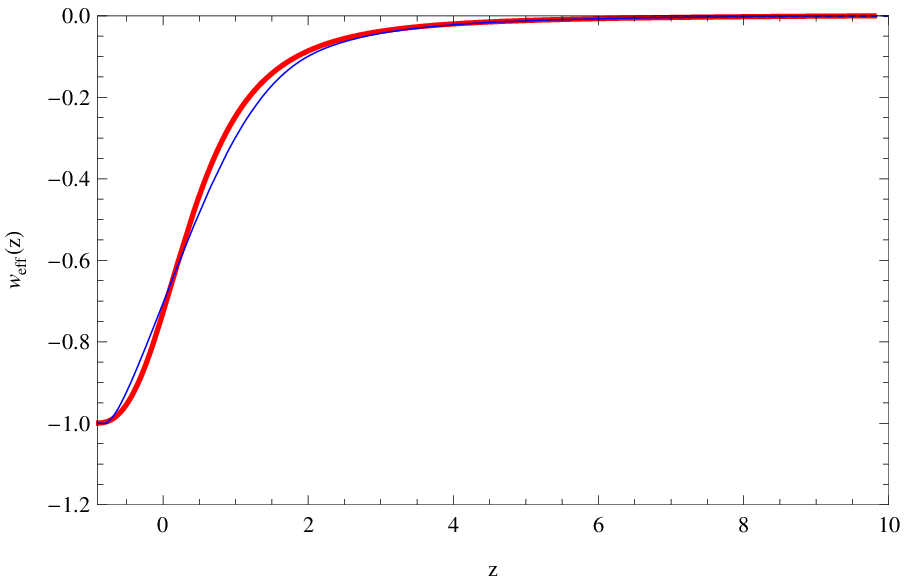}
\caption{Comparison of the dark energy equation of state parameter $\omega_{DE}(z)$ over z (right plot), of the scaled dark energy density $y_H(z)=\frac{\rho_{DE}}{\rho_m^{(0)}}$ over z (left plot), and of the effective equation of state parameter $\omega_{eff}(z)$ over z  (bottom plot)  for the exponential $F(R)$ model of Eq. (\ref{expmodnocurv}). The mimetic $F(R)$ case corresponds to the red curves , while the blue curves correspond to the exponential with power-law term ordinary $F(R)$ model.} \label{plot2}
\end{figure}
In Fig. \ref{plot2} we compare the scaled dark energy density $y_H(z)$ (left plot), the dark energy equation of state parameter $\omega_{DE}(z)$ (right plot) and the effective equation of state parameter $\omega_{eff}(z)$ for the ordinary $F(R)$ and the mimetic $F(R)$ gravity. The red curves correspond to the mimetic $F(R)$ model while the blue curves correspond to the ordinary $F(R)$ model. By looking at Fig. \ref{plot2} it can be seen that the differences are traced in the scaled dark energy density $y_H(z)$ and in the dark energy equation of state parameter $\Omega_{DE}(z)$ and this can be explained easily since the higher derivatives of the Hubble rate are involved in these two physical quantities. The mimetic $F(R)$ case has the appealing feature that the dark energy oscillations are damped in comparison to the ordinary $F(R)$ exponential model and more importantly, no curvature corrections are needed in the mimetic $F(R)$ case in order to make the oscillations damped.
\begin{table*}
\small
\caption{\label{tablei} The dark energy equation of state parameter $\omega_{DE}(z)$ and the effective equation of state parameter $\omega_{eff}$, at $z=0$ (present time) for the mimetic exponential $F(R)$ model and for the ordinary exponential $F(R)$.}
\begin{tabular}{@{}crrrrrrrrrrr@{}}
\tableline
\tableline
\tableline
Model &  $\omega_{DE}(0)$ &  $\omega_{eff}(0)$
\\\tableline
Ordinary Exponential $F(R)$ Model &  $0.733292$ & $-0.936629$
 \\\tableline
 Mimetic Exponential $F(R)$ Model &  $0.704561$ & $-1.0029$
\\\tableline
Observational Data &  $0.721\pm 0.015$ & $-0.972\pm 0.06$
\\\tableline
\tableline
 \end{tabular}
\end{table*}
In addition, the total equation of state parameter $\omega_{eff}$ is quite similar in the two cases, but in order to have a clear picture of what do the models predict for present time, in Table \ref{tablei} we present the values of the effective equation of state parameter $\omega_{eff}(z)$ and of the dark energy density $\Omega_{DE}(z)$ at redshift $z=0$ which corresponds to present day. As it can be seen, the resulting values are very close to the observational data, but not in full agreement and the resulting picture is similar to the ordinary $F(R)$ gravity model. Now the question is how the $F(R)$ gravity model actually affects the late-time behavior and to this end in the next section we shall repeat the numerical analysis for a viable power-law model.

\subsection{The Power-law $F(R)$ Gravity Case}

\begin{figure}[h]
\centering
\includegraphics[width=15pc]{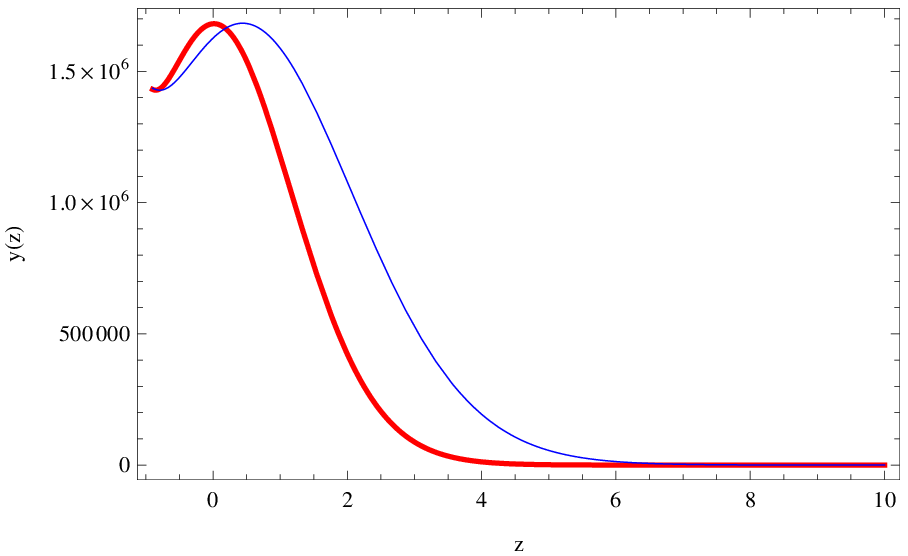}
\includegraphics[width=15pc]{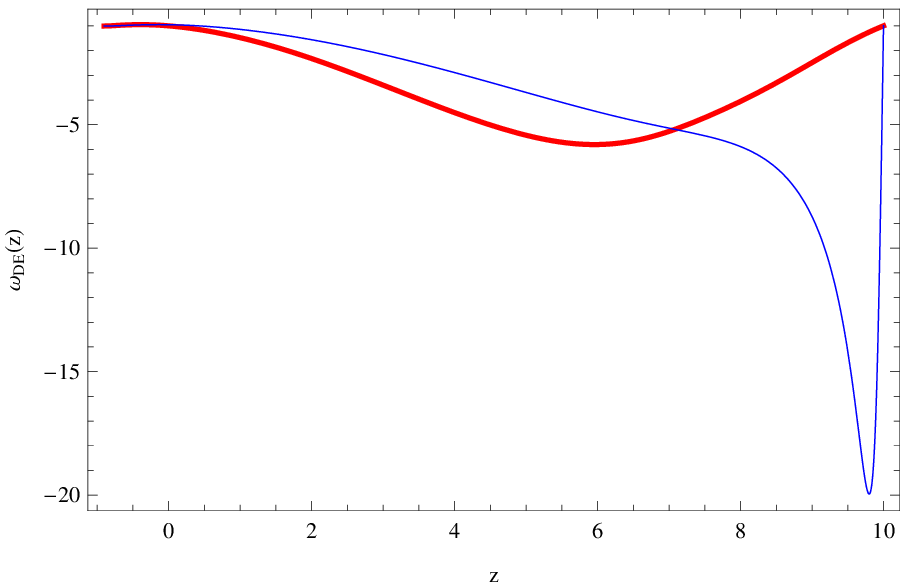}
\includegraphics[width=15pc]{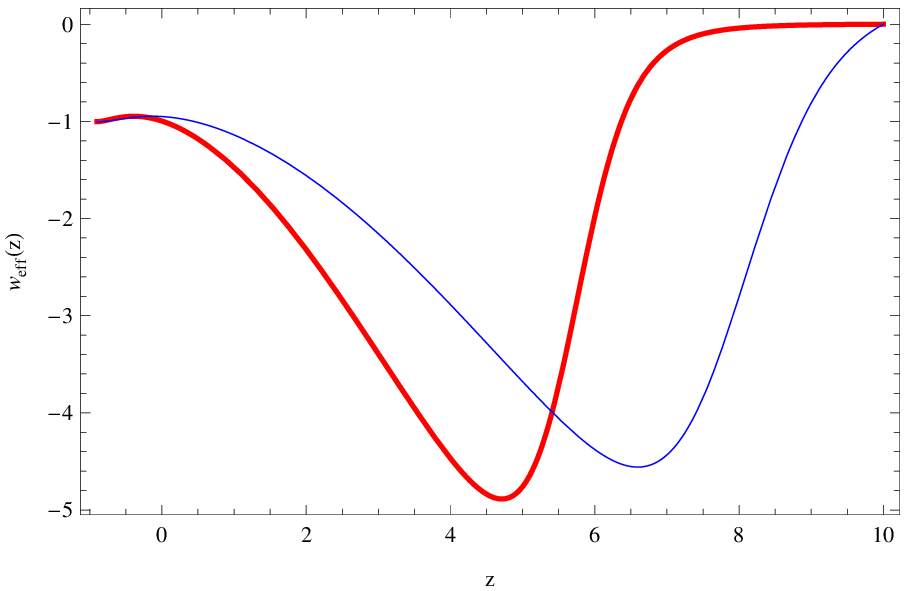}
\caption{Comparison of the dark energy equation of state parameter $\omega_{DE}(z)$ over z (right plot), of the scaled dark energy density $y_H(z)=\frac{\rho_{DE}}{\rho_m^{(0)}}$ over z (left plot), and of the effective equation of state parameter $\omega_{eff}(z)$ over z  (bottom plot)  for the power-law $F(R)$ model of Eq. (\ref{powerlaw}). The mimetic $F(R)$ case corresponds to the red curves, while the blue curves correspond to the power-law ordinary $F(R)$ model.} \label{plot4}
\end{figure}

Having the results of the previous section in mind it is worth studying another $F(R)$ gravity model, in order to see if the functional form of the $F(R)$ gravity alters the qualitative picture. As we demonstrate, the form of the $F(R)$ gravity indeed affects the late-time evolution and it is remarkable that the $F(R)$ gravities which describe a unification cosmology, generate a very smooth late-time evolution. Let us consider the model (\ref{powerlaw}), and we numerically solve the differential equation (\ref{mastereqn}) for the initial conditions and the values of the parameters that we considered in the previous section. In addition, for the ordinary $F(R)$ case, the $F(R)$ gravity has the following form,
\begin{equation}\label{powerlawcorrterm}
F(R)=R-\mu R^{n}-\tilde{\gamma}\Lambda \left (\frac{R}{3\tilde{m}^2} \right )^{1/3}\, ,
\end{equation}
so it is obvious by comparing to Eq. (\ref{powerlaw}), that a correction term is needed in the ordinary $F(R)$ case too. Following the procedure of the previous section, the function $\mathcal{Q}(z)$ at leading order, is in this case,
\begin{align}\label{qzpowerlaw}
& \mathcal{Q}(z)=\frac{3^n 4^{-1+n} \left(H_L^2\right)^n \left(3 H_I^3 n^2 (1+z)^{\frac{3}{H_L t_0}}+48 H_I^3 H_L^2 n (-1+2 n) t_0^2 (1+z)^{\frac{3}{H_L t_0}}+192 H_I^3 H_L^3 n t_0^3 (1+z)^{\frac{3}{H_L t_0}}\right) \mu }{H_L t_0^5 \kappa ^2}\\ \notag &
+\frac{3^n 4^{-1+n} \left(H_L^2\right)^n \left(H_L \left(-12 H_I^3 n (-1+2 n) t_0 (1+z)^{\frac{3}{H_L t_0}}-(-1+n) t_0^5 \left(-2+(1+z)^{\frac{6}{H_L t_0}}\right)\right)\right) \mu }{H_L t_0^5 \kappa ^2}\, .
\end{align}
The results of the numerical analysis appear in Fig. \ref{plot4}, where the red curves correspond to the mimetic $F(R)$ case, while the blue to the curvature corrected ordinary $F(R)$. As it can be seen, the situation is in general worse in comparison to the exponential model, since the amplitude of the oscillations is significantly larger for both the mimetic and ordinary $F(R)$. Moreover, in Table \ref{tableii} we present the values of the total equation of state parameter and of the dark energy equation of state parameter at present time. As it can be seen, the mimetic and ordinary $F(R)$ power law model have the same present time dark energy equation of state, but the total equation of state parameter is different. Also the resulting values are far from the observed values for the dark energy equation of state.
 In addition, the behavior of the scaled dark energy and of the total effective equation of state is similar in the mimetic and ordinary $F(R)$, while the dark energy equation of state is different.
 \begin{table*}
\small
\caption{\label{tableii} The values of the dark energy equation of state parameter $\omega_{DE}(z)$ and of the effective equation of state parameter $\omega_{eff}$, at redshift $z=0$ for the power law mimetic $F(R)$ model and for the ordinary power-law $F(R)$ model.}
\begin{tabular}{@{}crrrrrrrrrrr@{}}
\tableline
\tableline
\tableline
Model &  $\omega_{DE}(0)$ &  $\omega_{eff}(0)$
\\\tableline
Ordinary Power-law $F(R)$ Model &  $0.999999$ & $-0.950045$
 \\\tableline
Mimetic Power-law $F(R)$ Model  &  $0.9999$ & $-0.99642$
\\\tableline
Observational Data &  $0.721\pm 0.015$ & $-0.972\pm 0.06$
\\\tableline
\tableline
 \end{tabular}
\end{table*}
Let us here recapitulate what we obtained so far. Firstly let us note that the late-time evolution is strongly affected by the choice of the $F(R)$ gravity model. It is intriguing that for the exponential model which describes in a unified way the late and early-time acceleration, the late-time evolution has appealing properties, especially in the mimetic case. Also, for the exponential model, there is no need for curvature corrections in order to decrease the amplitude of the dark energy oscillations, since these are damped due to the presence of the mimetic potential. Therefore, the presence of the mimetic potential (\ref{leadingordpotexp}) and of the Lagrange multiplier (\ref{lambdaleading}) has a crucial effect on the late-time evolution.

\section{Effective Gravitational Constant and Subhorizon Growth Index for Exponential Mimetic $F(R)$ Gravity}

Now we turn our focus on two other physical quantities, namely the effective gravitational constant and also the growth index. The effective gravitational constant depends on the comoving wavenumber of the curvature perturbation, and it is calculated in the so-called subhorizon approximation. Particularly, the subhorizon approximation assumes that the wavenumber $k$ of the primordial matter density perturbation satisfies $\frac{k^2}{a^2}\gg H^2$, where $a$ and $H$ are the scale factor and the Hubble rate respectively. Note that the primordial modes evolve and become relevant at present day, so in some sense by calculating the effective gravitational constant at low redshifts, we find the effect of these subhorizon matter perturbation modes in the cosmological evolution of low redshifts.

\begin{figure}[h]
\centering
\includegraphics[width=15pc]{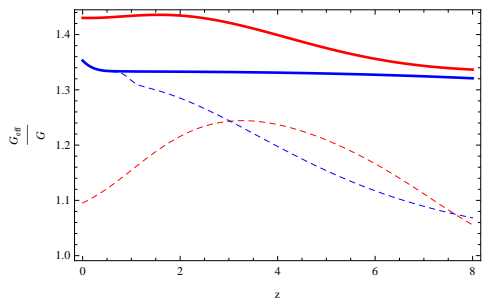}
\caption{The effective gravitational constant $G_{eff}$ over the Newton's constant $G$ for the mimetic $F(R)$ exponential gravity (red curves) and also for the ordinary $F(R)$ models (blue curves). The red thick line corresponds to $k=0.1$Mpc$^{-1}$ and the red dashed line to $k=0.0001$Mpc$^{-1}$, while the blue thick to $k=1$Mpc$^{-1}$ and the blue dashed to $k=0.1$Mpc$^{-1}$} \label{plotgeff}
\end{figure}

\begin{figure}[h]
\centering
\includegraphics[width=15pc]{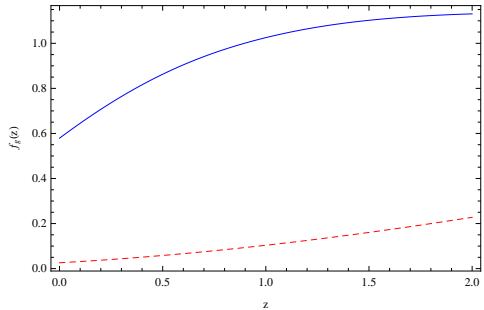}
\caption{The growth factor $f_g(z)$ over z, for $k=0.1$Mpc$^{-1}$ for the mimetic exponential $F(R)$ model (red curve) and for the ordinary $F(R)$ model (blue curve).} \label{plot6}
\end{figure}

The effective gravitational constant $G_{eff}$ of an $F(R)$ theory of gravity is given by \cite{exp1,dobado},
\begin{equation}\label{geff}
G_{eff}(a,k)=\frac{G}{8\pi F'(R)}\Big{[}1+\frac{\frac{k^2}{a^2}\frac{F''(R)}{F'(R)}}{1+3\frac{k^2}{a^2}\frac{F''(R)}{F'(R)}} \Big{]}\, .
\end{equation}
For the mimetic $F(R)$ case, the effective gravitational constant is also given by relation (\ref{geff}) as we now evince. The mimetic $F(R)$ theory is a theory of the form $F(R,\phi, X)$, where $X=-g^{\mu \nu}\partial_{\mu}\phi\partial_{\nu}\phi $, and for these theories the effective gravitational constant is of the form \cite{tsuji1},
\begin{equation}\label{correctgef}
G_{eff}(a,k)=\frac{G}{8\pi F_R}\frac{F_{X}+4\left(F_X \frac{k^2}{a^2}\frac{F_{RR}}{F_R}+\frac{F_{R\phi}}{F_R}\right)}{F_{X}+3\left(F_X \frac{k^2}{a^2}\frac{F_{RR}}{F_R}+\frac{F_{R\phi}}{F_R}\right)}\, ,
\end{equation}
with $F_R$ being the derivative of $F(R,\phi, X)$ with respect to the Ricci scalar $R$, $F_X$ with respect to $X$ and so on. Owing to the fact that $\partial_{\phi} F(R)=0$, the expression of Eq. (\ref{correctgef}) becomes identical to that of Eq. (\ref{geff}). In order to reveal the behavior of the effective gravitational constant we need to specify for which values of the wavenumber $k$ the subhorizon approximation is valid. We focus on the exponential $F(R)$ models we studied in the previous sections and for these models, the allowed values of the wavenumber are given in Table \ref{tableiiiv}.
\begin{table*}[h]
\small
\caption{\label{tableiiiv} Allowed values of the comoving wavenumber $k$ for the exponential mimetic and ordinary $F(R)$ models}
\begin{tabular}{@{}crrrrrrrrrrr@{}}
\tableline
\tableline
\tableline
Model & Allowed value of $k$
\\\tableline
Ordinary Exponential $F(R)$ Model &  $k>0.000115629$
 \\\tableline
Mimetic Exponential $F(R)$ Model  &  $k>0.0000934945$
 \\\tableline
 \end{tabular}
\end{table*}
In Fig. \ref{plotgeff} we plot the $z$-dependence of the effective gravitational constant over the Newton's constant $G$, for the mimetic $F(R)$ exponential gravity (red curves) and also for the ordinary $F(R)$ models (blue curves). The red thick line corresponds to $k=0.1$Mpc$^{-1}$ and the red dashed line to $k=0.0001$Mpc$^{-1}$, while the blue thick to $k=1$Mpc$^{-1}$ and the blue dashed to $k=0.1$Mpc$^{-1}$. As it can be seen, there are some differences between the mimetic and non-mimetic cases, but nevertheless, the predicted values of the fraction $G_{eff}/G$ are not very far from each other.
Having discussed the behavior of the effective gravitational constant, we proceed to the study of the growth index. The matter density perturbations in the mimetic $F(R)$ gravity are quantified in terms of the parameter $\delta =\frac{\delta \varepsilon_m}{\varepsilon_m}$, where $\epsilon_m\sim a^{-4}$, and notice that the mimetic potential and Lagrange multiplier affect the evolution of this parameter via the resulting Hubble rate, which depends on the scaled dark energy parameter $y_H(z)$ of the previous section. In addition, we need to note that the evolution of the Universe is by no means radiation dominated, owing to the existence of the potential and the Lagrange multiplier. For example in the mimetic $F(R)$ case without potential this is quantified by the presence of a term $C/a^3$ in the gravitational equations, and also in the case at hand, the presence of the potential itself manifests  that the evolution is not radiation dominated.  The parameter $\delta$ satisfies the following differential equation,
\begin{equation}\label{matterperturb}
\ddot{\delta}+2H\dot{\delta}-4\pi G_{eff}(a,k)\varepsilon_m\delta =0\, ,
\end{equation}
but we shall rewrite this equation in terms of the growth factor $f_g(z)=\frac{\mathrm{d}\ln \delta}{\mathrm{d}\ln a}$, and the resulting equation is,
\begin{equation}\label{presenceofcoll}
\frac{\mathrm{d}f_g(z)}{\mathrm{d}z}+\Big{(}\frac{1+z}{H(z)}\frac{\mathrm{d}H(z)}{\mathrm{d}z}-2-f_g(z)\Big{)}\frac{f_g(z)}{1+z}+\frac{4\pi}{G}\frac{G_{eff}(a(z),k)}{(z+1)H^2(z)}\varepsilon_m=0
\end{equation}
By solving numerically the equation (\ref{presenceofcoll}), using the same numerical values for the parameters we used in the previous sections and also by using the initial condition $f_g(z_{fin}, k)=1$, with $z_{fin}=10$, we obtain the results which appear in Fig. \ref{plot6}. The red curve corresponds to the mimetic exponential $F(R)$ model while the blue curve corresponds to the ordinary exponential $F(R)$ model. As it can be seen, the late-time evolution of the two models can be distinguished by studying the growth factor, since there is a significant difference between the cases at hand.

\section{Conclusions}

In this paper we demonstrated how the $F(R)$ gravity, the mimetic potential and the Lagrange multiplier affect the late-time cosmological evolution. We thoroughly investigated the late-time era of the mimetic $F(R)$ gravity and we compared with the ordinary $F(R)$ gravity case. As we showed, the late-time era is mainly controlled by the functional form of the $F(R)$ gravity and we illustrated this by studying two models, an exponential and a power-law model, which in the literature are considered viable. The results of the exponential model are in better agreement with the observations and the late-time behavior appears to be in general more ``smooth'', since the dark energy oscillations are damped. Also a major advantage of the mimetic approach is that the $F(R)$ gravity needs not to be altered by hand by adding curvature corrections, in order to make the amplitude of the oscillations smaller. This result however is model dependent and a deeper analysis is needed towards this research line. With regard to the late-time evolution, we quantified our analysis by studying three physical quantities, namely, the scaled dark energy, the dark energy equation of state and the total effective equation of state. As we demonstrated, for the exponential model the mimetic $F(R)$ gravity has smoother behavior in comparison to the ordinary $F(R)$ when the scaled dark energy and the dark energy equation of state are considered, however, the total effective equation of state behaves in a similar way in both cases. Significant differences between the ordinary $F(R)$ and the mimetic $F(R)$ cases occur when the growth factor is considered, at least for the exponential unification cosmology. Also the effective gravitational constant behaves in a different way in the two cases, but the differences are not significant.

An interesting task we did not address in this paper is the possibility of the phantom divide crossing after the end of the matter domination era and until the near or even the far future. Usually when this occurs, a Big Rip singularity or similar occurs, so this issue must be carefully addressed in a more focused work. The vital feature of this analysis would be the presence of the potential and Lagrange multiplier, so the question would be how to introduce a compensating dark fluid that would make the phantom crossing to disappear from the evolution. This task however exceeds the purposes of this paper and it is highly non-trivial to address it, so we defer this to a future work.

\section*{Acknowledgments}

This work is supported by Min. of Education and Science of Russia (V.K.O).

\end{document}